\documentclass[aps,pra,notitlepage,twocolumn,10pt,longbibliography]{revtex4-1}
\usepackage{graphicx}
\usepackage{bm}
\usepackage[colorlinks=true]{hyperref}
\usepackage{xcolor}
\usepackage{amsmath} 
\newcommand{\indices}[2]{_{#1}^{\phantom{#1}#2}}

\begin{document}

\title{Energy-momentum tensor of a ferromagnet}

\author{Sayak Dasgupta}
\author{Oleg Tchernyshyov}
\affiliation{
Department of Physics and Astronomy, 
Johns Hopkins University,
Baltimore, Maryland 21218, USA
}
\date{\today}

\begin{abstract}
The energy-momentum tensor of a ferromagnet derived according to the standard prescription of Noether's theorem has a major flaw: the term originating from the spin Berry phase is gauge-dependent. As a consequence, some physical quantities computed from the tensor show unphysical behavior. For example, the presence of a spin-polarized current does not affect the energy of the domain wall in the commonly accepted gauge, which implies---incorrectly---the absence of the adiabatic spin torque. In other gauges, the spin torque shows unphysical glitches occurring when the plane of magnetization crosses the Dirac string associated with a magnetic monopole in spin space. We derive a gauge-invariant energy-momentum tensor that is free from these artifacts but requires the addition of an extra spatial dimension, with the ferromagnet living on its boundary. It can be obtained most directly from the Wess-Zumino action for spins, which relies on the same extra dimension.
\end{abstract}

\maketitle

\section{Introduction}
\label{sec-intro}

Micromagnetics \cite{Aharoni1996} is a field theory of the ferromagnet. It treats magnetization $\mathbf M(t,\mathbf r)$ as a field slowly varying in both time and space. Although this approach cannot be applied on the atomic scale, it proved to be useful for understanding the structure and dynamics of magnetic solitons such as domain walls, vortices, and skyrmions, whose characteristic length scales are typically much longer than the atomic lattice constant. The theory is often stated in the form of the Landau-Lifshitz equation for the magnetization field. Its basic version, excluding the effects of dissipation, reads
\begin{equation}
\partial_t \mathbf m = 
	- \gamma \mathbf H_\mathrm{eff} \times \mathbf m 
    - (\mathbf u \cdot \nabla) \mathbf m.
\label{eq:LLG}
\end{equation}
Here $\mathbf m(t,\mathbf r) = \mathbf M(t,\mathbf r)/M$ is the unit vector parallel to magnetization and $\gamma$ is the gyromagnetic ratio. The effective field
\begin{equation}
\mathbf H_\mathrm{eff}(\mathbf r) 
	= - \frac{1}{M} 
    \frac{\delta U[\mathbf m]}{\delta \mathbf m(\mathbf r)}.
\label{eq:Heff}
\end{equation}
contains information about magnetic interactions and is derived from an energy functional $U[\mathbf m]$. The last term in Eq.~(\ref{eq:LLG}), first derived by \textcite{Bazaliy1998}, represents the adiabatic spin-transfer torque exerted by spin-polarized electric current. The drift velocity $\mathbf u$ is proportional to the current density and spin polarization \cite{Tatara2004, PhysRep.468.213, Brataas2017}. 

Even for the simplest forms of the energy functional $U[\mathbf m]$, the Landau-Lifshitz equation is a nonlinear partial differential equation. A full analytical solution is rarely possible. The difficulty of finding exact solutions necessitates the development of alternative approaches such as the use of conservation laws arising from symmetries. The symmetry of translations in space and time yields global conservation laws of momentum $\mathbf P$ and energy $E$ and a local conservation law, 
\begin{equation}
\partial_\beta T\indices{\alpha}{\beta} = 0,
\label{eq:T-conservation}
\end{equation}
for the energy-momentum tensor $T\indices{\alpha}{\beta}$. (Summation is implied over doubly repeated indices.) Conserved momenta have been used to study interaction of domain walls with with magnons \cite{Yan2013}, with spin-polarized current \cite{Chauleau2010}, and with one another \cite{Kosevich1977,Long1979}, the motion of skyrmions under an external force \cite{Komineas1996}, collisions of vortices \cite{Kovalev2002}, and scattering of spin waves on a skyrmion \cite{Iwasaki2014, Schutte2014}. 

It may be surprising that these conservation laws have not been used more widely in the studies of ferromagnetic solitons. The main obstacle is the treatment of the gyroscopic force associated with the precessional dynamics of spins in a ferromagnet. The standard derivation of the energy-momentum tensor through the application of Noether's theorem sometimes yields unphysical answers. 

To expose the nature of the problem, we rewrite the Landau-Lifshitz equation (\ref{eq:LLG}) in a modified form \cite{transverse}, 
\begin{equation}
\mathcal J \mathbf m \times 
	(\partial_t + \mathbf u \cdot \nabla) \mathbf m
- \frac{\delta U[\mathbf m]}{\delta \mathbf m}
= 0.
\label{eq:LLG-mod}
\end{equation}
Here $\mathcal J = \mathcal M/\gamma$ is the density of angular momentum. Equation (\ref{eq:LLG-mod}) expresses the balance of forces acting on a particle confined to move on a sphere of radius $|\mathbf m| = 1$. The first term $\mathcal J \mathbf m \times \partial_t \mathbf m$ is proportional to the velocity $\partial_t \mathbf m$ and is perpendicular to it, thus resembling a Lorentz force acting on an electric charge. The ``magnetic field'' in this analogy points in the radial direction, $\mathbf b = - \mathcal J \mathbf m$, as if it were created by a magnetic monopole of strength $\mathcal J$ at the center of the sphere. In the Lagrangian formulation, the Lorentz force is encoded through a gauge potential $\mathbf a(\mathbf m)$, whose curl yields the magnetic field, 
\begin{equation}
\nabla_{\mathbf m} \times \mathbf a(\mathbf m) = - \mathcal J \mathbf m.
\label{eq:curl-a}
\end{equation}

Finding a solution of Eq.~(\ref{eq:curl-a}) runs into two problems. First, it is not possible to find $\mathbf a(\mathbf m)$ that is well defined on the entire sphere. The following axially symmetric solutions \cite{Haldane1986} have a singularity at $\mathbf m = \mathbf m_s$: 
\begin{equation}
\mathbf a(\mathbf m)
	= \mathcal J 
		\frac{\mathbf m_s \times \mathbf m}
        	{1 - \mathbf m_s \cdot \mathbf m}. 
\label{eq:gauge-potential-axial}
\end{equation}
The singularity is the location of a Dirac string carrying the magnetic flux $+4 \pi \mathcal J$ that compensates the net flux $- 4 \pi \mathcal J$ of the magnetic monopole. Second, gauge potentials are not uniquely defined. Any gauge transformation 
\begin{equation}
\mathbf a(\mathbf m) \mapsto \mathbf a(\mathbf m) + \nabla_{\mathbf m} \chi(\mathbf m),
\label{eq:gauge-transform}
\end{equation}
where $\chi$ is a smooth function of $\mathbf m$, leaves the magnetic field $\mathbf b(\mathbf m)$ unchanged.  

The Lagrangian that yields the equations of motion (\ref{eq:LLG}) and (\ref{eq:LLG-mod}) is
\begin{equation}
L = \int dV \, \mathbf a(\mathbf m) \cdot
	u^\alpha \partial_\alpha \mathbf m
    - U[\mathbf m],
\label{eq:L}
\end{equation}
where we introduced a relativistic shorthand notation $u^\alpha \partial_\alpha = \partial_t + \mathbf u \cdot \nabla$ with index $\alpha = 0$ for time and $\alpha \geq 1$ for space dimensions; $u^\alpha = (1,\mathbf u)$, and $\partial_\alpha = (\partial_t, \nabla)$. We use the Minkowski metric $\eta_{\alpha\beta} = \mathop{\mathrm{diag}}{(1,-1,-1,-1)}$ in 3 spatial dimensions. 

The energy-momentum tensor can now be obtained in the standard way from the Lagrangian \cite{PeskinSchroeder}: 
\begin{eqnarray}
T\indices{\alpha}{\beta} 
    &=& \frac{\partial \mathcal L}
    {\partial (\partial_\beta \mathbf m)}
    \cdot \partial_\alpha \mathbf m
    - \delta_\alpha^\beta \, \mathcal L
\nonumber\\
    &=&
	\left(
		\delta_\alpha^\nu \delta_\mu^\beta
        - \delta_\alpha^\beta \delta_\mu^\nu
    \right)
	u^\mu \, 
    \mathbf a(\mathbf m) \cdot \partial_\nu \mathbf m
    + \ldots
\label{eq:T}
\end{eqnarray}
The omitted terms come from the potential energy functional $U[\mathbf m]$. The energy-momentum tensor (\ref{eq:T}) is gauge-dependent and changes under a gauge transformation (\ref{eq:gauge-transform}) as follows: 
\begin{equation}
T\indices{\alpha}{\beta} 
	\mapsto T\indices{\alpha}{\beta} 
    + \left(
		\delta_\alpha^\nu \delta_\mu^\beta
        - \delta_\alpha^\beta \delta_\mu^\nu
    \right) u^\mu \partial_\nu \chi.
\end{equation}
The tensor components are gauge dependent and therefore unphysical. For example, energy density transforms as $T\indices{0}{0} \mapsto T\indices{0}{0} - u^{i}\partial_{i}\chi$, where Roman indices denote spatial coordinates. However, the divergence of the tensor $\partial_\beta T\indices{\alpha}{\beta}$ is not affected by gauge transformations, provided that the 4-velocity $u$ is constant in spacetime, $\partial_\beta u^\mu = 0$. Thus the local conservation law (\ref{eq:T-conservation}) is a physical statement. 

A somewhat similar problem is encountered in the derivation of the energy-momentum tensor of the electromagnetic field. A direct application of Noether's theorem to the Lagrangian density $\mathcal L = - F_{\mu\nu}F^{\mu\nu}/16\pi$ yields a gauge-dependent expression 
\begin{equation}
T\indices{\alpha}{\beta} = - \frac{F^{\beta\gamma}\partial_\alpha A_\gamma}{4\pi} 
+ \delta_\alpha^\beta \, 
\frac{F_{\mu\nu}F^{\mu\nu}}{16\pi}.
\end{equation}
The gauge dependence is removed by a transformation 
\begin{equation}
{T_\alpha}^\beta \mapsto {T_\alpha}^\beta + \partial_\gamma {\Sigma_\alpha}^{\beta\gamma},
\label{eq:T-transform}
\end{equation}
which does not spoil the local conservation law (\ref{eq:T-conservation}) if  $\Sigma\indices{\alpha}{\beta\gamma}$ is antisymmetric under the exchange of its upper indices \cite{PeskinSchroeder}. Choosing $\Sigma\indices{\alpha}{\beta\gamma} = A_\alpha F^{\beta\gamma}/4\pi$ yields the familiar gauge-invariant energy-momentum tensor of the electromagnetic field
\begin{equation}
T\indices{\alpha}{\beta} = - \frac{F_{\alpha\gamma}F^{\beta\gamma}}{4\pi}
+ \delta_\alpha^\beta \, 
\frac{F_{\mu\nu}F^{\mu\nu}}{16\pi}.
\end{equation}

Our attempts to find such a transformation for the stress-energy tensor of a ferromagnet (\ref{eq:T}) were unsuccessful. Instead, we relied on our previous work \cite{Tchernyshyov2015}, which resolved a similar problem for a related global quantity: the linear momentum $P_i$ of a ferromagnetic soliton. Because it is directly related to the $T\indices{i}{0}$ component of the energy-momentum tensor, we have been able to extend the recipe for constructing gauge-invariant conserved momenta to the components of the stress-energy tensor $T\indices{\alpha}{\beta}$. 

Gauge invariance of the energy-momentum tensor is achieved at the price of adding an extra dimension to the usual space and time. In a sense, the $(d+1)$-dimensional spacetime is viewed as the boundary of a $(d+2)$-dimensional manifold. A similar extension is made in the construction of the Wess-Zumino action \cite{Volovik1987, Fradkin2013, Abanov2017}. In fact, the gauge-invariant energy-momentum tensor can be most directly obtained from the Wess-Zumino action for the ferromagnet. 

The rest of the paper is organized as follows. In Sec.~\ref{sec:T-canonical}, we illustrate the problematic nature of the canonical energy-momentum tensor (\ref{eq:T}) on the examples of a domain wall in $d=1$ and of a vortex in $d=2$. In Sec.~\ref{sec:T-gauge-invariant}, we derive a gauge-invariant version of this quantity and show on the same examples that it provides a more sensible alternative. Single-valuedness of the physical quantities obtained from this tensor is discussed in Sec.~\ref{sec:single-valuedness}. Concluding remarks are made in Sec.~\ref{sec:discussion}. The Wess-Zumino action for a single spin is reviewed in the Appendix.

\section{Canonical energy-momentum tensor}
\label{sec:T-canonical}
In this section we highlight the issues that arise from using the standard energy-momentum tensor Eq.(\ref{eq:T}) to calculate the adiabatic spin torque on a domain wall and the force on a ferromagnetic vortex. 

\subsection{Domain wall in one spatial dimension}
\label{sec:T-canonical-domain-wall}

A sample energy functional for an easy-axis ferromagnet in one spatial dimension reads 
\begin{equation}
U[\mathbf m(x)] = \int dx 
	\left[
    	A \frac{(\partial_x \mathbf m)^2}{2} 
        + K \frac{(\mathbf m \times \mathbf e_3)^2}{2}
    \right],
\label{eq:U-easy-axis}
\end{equation}
where $A>0$ is the strength of exchange interactions (favoring a uniform magnetization, $\partial_x \mathbf m = 0$), $K>0$ is the anisotropy constant, and $\mathbf e_3 = (0,0,1)$ is a unit vector along the easy direction. A solution with a static domain wall separating domains with $\mathbf m = - \mathbf e_3$ and $+ \mathbf e_3$ can be found as a local minimizer of the energy (\ref{eq:U-easy-axis}): 
\begin{equation}
\cos{\theta(x)} = \sigma \tanh{\frac{x-X}{\lambda}},
\quad
\phi(x) = \Phi.
\label{eq:domain-wall}
\end{equation}
Here $\theta$ and $\phi$ are the polar and azimuthal angles parameterizing the unit vector 
\begin{equation}
\mathbf m = 
    (\sin{\theta}\cos{\phi},
    \sin{\theta}\sin{\phi},
    \cos{\theta}),   
\end{equation}
$\lambda = \sqrt{A/K}$ is the width of the domain wall, $X$ and $\Phi$ are collective coordinates of the zero modes associated with translational and rotational symmetries of the energy (\ref{eq:U-easy-axis}), and $\sigma = \pm 1$ is a $Z_2$ topological charge of the domain wall. 

A spin-polarized electric current flowing with a drift velocity $u$ \cite{Tatara2004, PhysRep.468.213, Brataas2017} exerts on the domain wall a torque (a generalized force in the $\Phi$ channel) $F_\Phi = -2 \sigma \mathcal J u$.  We shall attempt to recover this torque by evaluating the energy of the domain wall 
\begin{equation}
E = \int dx \, T\indices{0}{0}
	= U 
	- u\int^{\infty}_{-\infty} dx \,
    	\mathbf{a}\cdot \partial_{x}\mathbf{m}. 
	= U + u P.
\label{eq:energy-domain-wall}
\end{equation}
Here $U = 2 \sqrt{AK}$ is the energy of a domain wall (\ref{eq:domain-wall}) in the absence of a spin current and 
\begin{equation}
P = - P_1 
	= - \int dx \, T\indices{1}{0} 
	= - \int^{\infty}_{-\infty} dx \,
    	\mathbf{a}\cdot \partial_{x}\mathbf{m}
\label{eq:P-canonical-domain-wall}
\end{equation}
is the (canonical) linear momentum. 

As discussed in detail in Ref.~\onlinecite{Tchernyshyov2015}, the canonical momentum (\ref{eq:P-canonical-domain-wall}) is a poorly defined quantity for a domain wall. Under a gauge transformation (\ref{eq:gauge-transform}), $P \mapsto P - \left. \chi(\mathbf m(x)) \right|_{-\infty}^{+\infty}$. Because $\mathbf m(+\infty) \neq \mathbf m(-\infty)$ for a domain wall, momentum $P$ is gauge-dependent and so is the energy $E = U + u P$. 

For the standard gauge choices (\ref{eq:gauge-potential-axial}) with the Dirac string at $\mathbf m_s = \pm \mathbf e_3$, 
\begin{equation}
P   = - \int dx \,
    	\mathcal J (\cos{\theta} \pm 1) \partial_x \phi 
    = 0
\end{equation}
because $\phi(x) = \Phi = \mathrm{const}$. Then the energy (\ref{eq:energy-domain-wall}) is independent of $\Phi$, implying---incorrectly---the absence of spin torque.

A related problem arises for the energy flux 
\begin{equation}
S = T\indices{0}{1} 
	= \mathcal J u (\cos{\theta} \pm 1)
	\, \partial_t \phi
	+ A \, \partial_t \mathbf m \cdot \partial_x \mathbf m.
\label{eq:S-domain-wall-canonical}
\end{equation}
For a rigidly rotating domain wall (\ref{eq:domain-wall}), both derivatives $\partial_t \mathbf m$ and $\partial_x \mathbf m$ vanish at spatial infinity, so the second term in the energy flux (\ref{eq:S-domain-wall-canonical}) related to exchange energy does not contribute. The spin-current term, however, remains finite at one of the ends. The total energy flowing out to infinity per unit time is $\left. S(x) \right|_{-\infty}^{+\infty} = 2 \sigma \mathcal J u$. This is clearly unphysical as there is no dynamics of magnetization, $\partial_t \mathbf m = 0$, far away from a domain wall. A finite energy outflow is an artifact. 

Both paradoxes can be traced to the singularity of the gauge potential $\mathbf a(\mathbf m)$ at one of the ground states of $\mathbf m = +\mathbf e_3$ or $-\mathbf e_3$. The vanishing of $\partial_t \mathbf m$ far away from the domain wall is compensated by the divergence of $\mathbf a$ on the Dirac string so that $S = u \, \mathbf a \cdot \partial_t \mathbf m$ remains finite. 

\begin{figure}
 \includegraphics[width=\columnwidth]{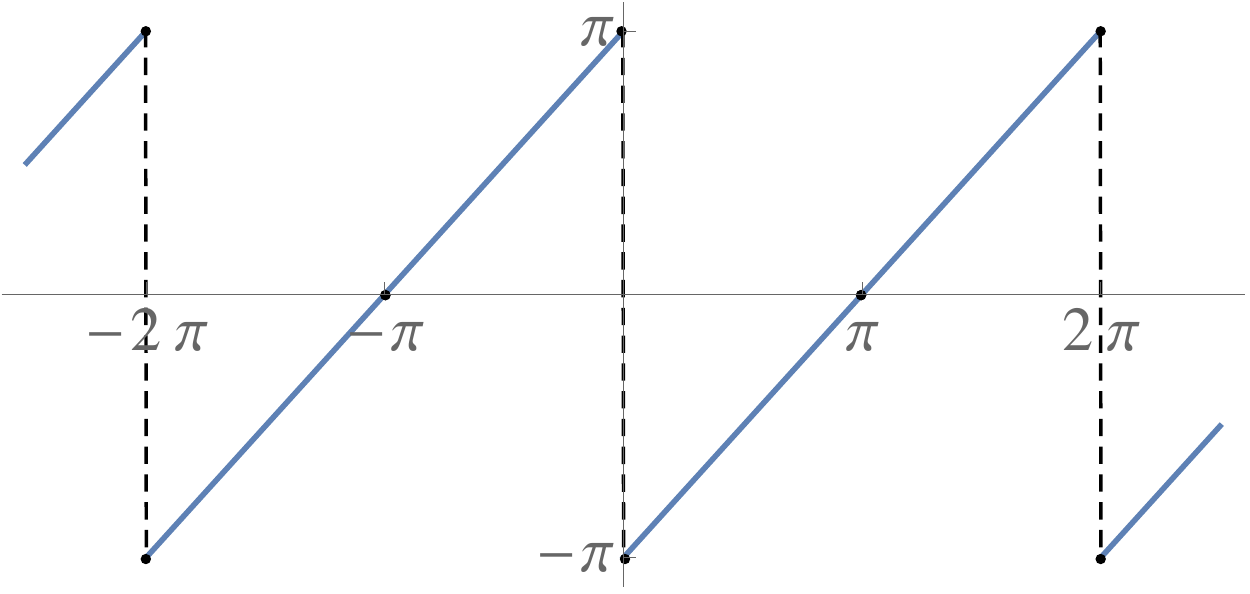}
\caption{Piecewise-linear function $f(\Phi) = - 2 \arctan{\cot{(\Phi/2)}}$.}
\label{fig:arctan-cot}
\end{figure}

The problem can be resolved in part by shifting the Dirac string away from ground states $\mathbf m = \pm \mathbf e_3$. Doing so would eliminate the unphysical energy flux far away from the domain wall. Canonical momentum is \cite{Tchernyshyov2015}
\begin{equation}
P = - 4 \sigma \mathcal J 
    	\arctan{\cot{\frac{\Phi - \phi_s}{2}}}.
\end{equation}
It is a piecewise-linear function of $\Phi$ with period $2\pi$  and with jump discontinuities at $\Phi = \phi_s + 2\pi n$, Fig.~\ref{fig:arctan-cot}. At these points, the line $\mathbf m(x)$ on the unit sphere crosses the Dirac string at $\mathbf m_s$. The energy $E = U + u P$ is now dependent on the azimuthal angle $\Phi$ and yields the spin torque 
\begin{equation}
F_\Phi = - \frac{\partial E}{\partial \Phi}
	= - 2 \sigma \mathcal J u 
    + 4\pi \sigma \mathcal J u 
    	\sum_n \delta(\Phi - \phi_s - 2\pi n).
\label{eq:torque-with-glitches}
\end{equation}
The result would be valid were it not for the singularities at $\Phi = \phi_s + 2\pi n$, occurring whenever the Dirac string happens to be in the plane of the domain wall. Similar glitches occur in the linear momentum of a ferromagnetic soliton \cite{Haldane1986}. These artifacts are unavoidable in the framework of a classical theory and require quantum mechanics to make the Dirac string invisible.

We have thus met with only partial success in defining the tensor of energy-momentum for a domain wall in the presence of adiabatic spin torque.  

\subsection{Vortex in two spatial dimensions}
\label{sec:T-canonical-vortex}

The simplest model of a thin-film ferromagnet with an easy-plane anisotropy is described by an energy functional similar to Eq.~(\ref{eq:U-easy-axis}), with two modifications: the spatial coordinates are now $\mathbf r = (x,y)$ and the anisotropy constant is now negative, $K<0$, which makes $\mathbf e_3$ the hard axis: 
\begin{equation}
U = \int dx \, dy 
	\left[
    	A \frac{(\partial_x \mathbf m)^2 
        	+ (\partial_y \mathbf m)^2}{2}
        + K \frac{(\mathbf m \times \mathbf e_3)^2}{2}
    \right].
\label{eq:energy-thin-film}
\end{equation}

A vortex configuration minimizing the energy (\ref{eq:energy-thin-film}) has the following profile for the azimuthal angle:
\begin{equation}
\phi (\mathbf r) 
	= n\arctan\left(\frac{y-Y}{x-X}\right) + \phi_0,
\label{eq:vortex}
\end{equation}
Here $\mathbf R = (X,Y)$ is the center of the vortex, $\phi_0$ is a constant, and 
\begin{equation}
n = \frac{1}{2\pi} \oint_C 
	d \mathbf r \cdot \mathbf \nabla \phi
\end{equation}
is an integer-valued topological charge inside contour $C$ known as the vortex number. In a circular region of characteristic size $\lambda = \sqrt{A/|K|}$, known as the vortex core, magnetization comes out of the easy plane to prevent the divergence of the exchange energy: 
\begin{equation}
\cos{\theta(\mathbf r)} = p f(|\mathbf r - \mathbf R|/\lambda).
\end{equation}
Here the function $f(\rho)$ smoothly interpolates from $f(0) = 1$ to $f(\infty) = 0$. Its precise form will not be needed for our purposes. The core polarity $p = \pm 1$ together with the vortex number $n$ determine another topological charge of the vortex, the skyrmion number  
\begin{equation}
Q = \frac{1}{4\pi}\int dx \, dy \, 
	\mathbf{m} \cdot 
        	(\partial_x \mathbf{m} \times \partial_y \mathbf{m})
	= \frac{np}{2}.
\label{eq:skyrmion-number}
\end{equation}
It is worth noting that contributions to the skyrmion number (\ref{eq:skyrmion-number}) are localized near the core; away from the core, magnetization lies in the easy plane so that the vectors $\mathbf m$, $\partial_x \mathbf{m}$, and $\partial_y \mathbf{m}$ are coplanar and the skyrmion density 
\begin{equation}
\rho = \frac{1}{4\pi}\mathbf m \cdot (\partial_x \mathbf{m} \times \partial_y \mathbf{m}) 
\label{eq:skyrmion-density}
\end{equation}
vanishes.

By working along the lines of Sec.~\ref{sec:T-canonical-domain-wall}, we obtain the energy of a vortex, 
\begin{equation}
E = U - u^i P_i = U + \mathbf u \cdot \mathbf P,
\label{eq:E-vortex}
\end{equation}
where now $\mathbf u = (u^x,u^y)$ and $\mathbf P = (P^x,P^y)$. Canonical momentum $\mathbf P$ can be computed in a standard gauge with the Dirac string away from magnetization at the core, $\mathbf m_s = - p \mathbf e_3$: 
\begin{equation}
P^i = - P_i = - \mathcal J \int dx \, dy \, 
	(\cos{\theta} - p) \partial_i \phi.
\end{equation}
The integrand decays very slowly with the distance $r$ from the vortex center (as $1/r$) and requires long-distance regularization. For a vortex in a disk of a finite radius, subtraction of the momentum of a reference state with the vortex at the center yields \cite{Tchernyshyov2015}
\begin{equation}
P^i(\mathbf R) - P^i(0) 
	= - \pi n p \mathcal J \epsilon_{ij}X^j.
\end{equation}
With the aid of Eqs.~(\ref{eq:skyrmion-number}) and (\ref{eq:E-vortex}), we find the spin-torque force on the vortex,
\begin{equation}
F^i = 2\pi Q \mathcal J \epsilon_{ij} u^j.
\label{eq:F-vortex-take-1}
\end{equation}
As we shall see below, this result is off by a factor of $1/2$. 

An alternative way to compute the force is via the stress tensor $\sigma^{ij} = - T\indices{i}{j}$, whose divergence yields the force density. For a stationary vortex, $\partial_t \mathbf m = 0$, the force density 
\begin{eqnarray}
f^i &=& \partial_j T\indices{i}{j}
	= u^j 
    (\partial_j \mathbf a \cdot \partial_i \mathbf m
    - \partial_i \mathbf a \cdot \partial_j \mathbf m)
\nonumber
\\
	&=& \mathcal J u^j \mathbf m \cdot
    	(\partial_i \mathbf m \times
        \partial_j \mathbf m)
\label{eq:force-density-vortex}
\end{eqnarray}
is proportional to the density of skyrmion charge (\ref{eq:skyrmion-density}), which quickly vanishes away from the vortex core. Integrating over the area yields the force 
\begin{equation}
F^i = \int dx \, dy \, f^i 
	= 4\pi Q \mathcal J \epsilon_{ij} u^j,
\label{eq:F-vortex-take-2}
\end{equation}
which disagrees with Eq.~(\ref{eq:F-vortex-take-1}) by a factor of 2.

\section{Gauge-invariant energy-momentum tensor}
\label{sec:T-gauge-invariant}

\subsection{Conjectured form of the tensor}

The inconsistencies noted in the previous section are rooted in the presence of a gauge potential in the Lagrangian (\ref{eq:L}). A natural way to resolve them is to obtain a gauge-invariant energy-momentum tensor. 

A similar paradox has been resolved recently for related quantities, conserved momenta of ferromagnetic solitons \cite{Tchernyshyov2015}.  The gauge-dependent canonical linear momentum  
\begin{equation}
P^i = - \int dV \, \mathbf a(\mathbf m) \cdot \partial_i \mathbf m
\end{equation}
is not a physical quantity and must be replaced with a gauge-invariant momentum defined as follows. Take a magnetization configuration $\mathbf m(t,\mathbf r)$ and make an infinitesimal deformation $\delta \mathbf m(t,\mathbf r)$. The resulting infinitesimal change in the linear momentum is 
\begin{equation}
\delta P^i = - \mathcal J \int dV \, \mathbf m \cdot \left( \partial_i \mathbf m \times \delta \mathbf m \right).
\label{eq:dP}
\end{equation}
By integrating the increment $\delta P^i$, we can obtain the finite difference of momenta $P^i_1 - P^i_0$ between any two configurations $\mathbf m_0(t,\mathbf r)$ and $\mathbf m_1(t,\mathbf r)$ continuously deformable into each other. To formalize this, add a continuous parameter $s$ so that $\mathbf m(t,\mathbf r,s)$ varies smoothly with $s$ and 
\begin{equation}
\mathbf m(t,\mathbf r,0) = \mathbf m_0(t,\mathbf r),
\quad
\mathbf m(t,\mathbf r,1) = \mathbf m_1(t,\mathbf r).
\label{eq:s-extension}
\end{equation}
Then \cite{Tchernyshyov2015}
\begin{equation}
P^i_1 - P^i_0
= - \mathcal J \int_0^1 ds \int dV \, \mathbf m \cdot \left( \partial_i \mathbf m \times \partial_s \mathbf m \right).
\label{eq:Delta-P}
\end{equation}
This expression was first obtained by \textcite{Thiele1976}. 

One might worry that the result of the integration in Eq.~(\ref{eq:Delta-P}) may depend on the path from $\mathbf m_0(t,\mathbf r)$ and $\mathbf m_1(t,\mathbf r)$ and then momentum difference $P^i_1 - P^i_0$ will be ill-defined (see, e.g., \textcite{Thiaville2007}). Fortunately, continuous deformations of the path do not affect the integral \cite{Tchernyshyov2015}. Therefore, the momentum difference (\ref{eq:Delta-P})  depends only on the topology of the path but not on its precise geometry. More on that in Sec.~\ref{sec:single-valuedness}.

As will become clear shortly, it is convenient to think of the parameter $s$ as of one more spatial dimension in a $(d+2)$-dimensional spacetime. Then the right-hand side of Eq.~(\ref{eq:Delta-P}) can be viewed as an integration over the now $(d+1)$-dimensional space $(\mathbf r,s)$. The integrand is the density of momentum in that space: 
\begin{equation}
T\indices{i}{0} = \mathcal J \mathbf m \cdot \left( \partial_i \mathbf m \times \partial_s \mathbf m \right).
\end{equation}
The other components of the energy-momentum tensor $T\indices{\alpha}{\beta}$ can be inferred by examining the index structure of Eq.~(\ref{eq:T}). We thus conjecture that the energy-momentum tensor in this $(d+2)$-dimensional spacetime is
\begin{equation}
T\indices{\alpha}{\beta} = 
    \mathcal{J}
    (\delta_\alpha^\nu \delta_\mu^\beta
        - \delta_\alpha^\beta \delta_\mu^\nu)
    \, u^{\mu} \, \mathbf{m} \cdot 
    (\partial_\nu \mathbf{m} 
    \times \partial_s \mathbf{m}) + \ldots,
\label{eq:T-extra-dim}
\end{equation}
with the omitted terms arising from the potential energy $U[\mathbf m]$. Eq.~(\ref{eq:T-extra-dim}) is the main result of this paper.

\subsection{Derivation from the spin Wess-Zumino action}

The gauge-invariant tensor (\ref{eq:T-extra-dim}) can be obtained directly from a gauge-invariant action for a ferromagnet. The standard action includes a gauge-dependent term $S = \int dt \int dV \, \mathbf a(\mathbf m) \cdot u^\alpha \partial_\alpha \mathbf m$. It can be made gauge-invariant at the expense of introducing an extra dimension $s$ \cite{Volovik1987, Fradkin2013, Abanov2017}: 
\begin{equation}
S_\text{WZ} 
= \int dt \int dV \int_0^1 ds \,
\mathcal J u^\alpha \mathbf m \cdot 
(\partial_\alpha \mathbf m 
\times \partial_s \mathbf m).
\label{eq:S-WZ}
\end{equation}

Although the Wess-Zumino action (\ref{eq:S-WZ}) includes an integral over a $(d+2)$-dimensional spacetime, its variation is nonzero only at the boundaries of the extra dimension, $s=0$ and $1$. These boundaries represent the physical $(d+1)$-dimensional spacetime. Each boundary can be thought of as an independent ferromagnet with its own spacetime. The net momentum of the system $P^i_1 - P^i_0$ and the net energy $E_1 - E_0$ come strictly from the boundaries. See Appendix \ref{app:WZ} for further details. 

Application of Noether's theorem to the Wess-Zumino action (\ref{eq:S-WZ}) immediately yields the energy-momentum tensor (\ref{eq:T-extra-dim}). The physical components of the tensor ($\alpha, \beta \neq s$) are not affected by the addition of the extra dimension $s$ because $u^s = 0$. 

Next we illustrate the application of the energy-momentum tensor (\ref{eq:T-extra-dim}) on the familiar examples of a domain wall in $d=1$ dimension and of a vortex in $d=2$ dimensions. 


\subsection{Domain wall in one spatial dimension}
\label{sec:T-gauge-invariant-domain-wall}

The energy difference between two magnetization configurations $\mathbf m_0(t,x)$ and $\mathbf m_1(t,x)$ living at the two boundaries of the 3-dimensional spacetime $(t,x,s)$ is obtained by integrating the energy $T\indices{0}{0}$ over the $(1+1)$-dimensional space $(x,s)$:
\begin{equation}
E_1 - E_0 = - \mathcal J u \int dx \int_0^1 ds \, \mathbf m 
    \cdot (\partial_x \mathbf m 
        \times \partial_s \mathbf m).
\label{eq:dE-domain-wall}
\end{equation}
For a rigid domain wall (\ref{eq:domain-wall}), 
\begin{equation}
\partial_s \mathbf m
    = \frac{\partial \mathbf m}{\partial X} \, \partial_s X
    + \frac{\partial \mathbf m}{\partial \Phi} \, \partial_s \Phi.
\end{equation} 
The first term does not contribute because $\partial \mathbf m /\partial X = - \partial_x \mathbf m$ so that the cross product in Eq.~(\ref{eq:dE-domain-wall}) vanishes. The second term yields
\begin{equation}
E_1 - E_0 =  
\mathcal J u \int dx \int_0^1 ds \, 
\partial_x \cos{\theta} \, \partial_s \Phi
= 2 \sigma \mathcal J u (\Phi_1 - \Phi_0).
\label{eq:E-domain-wall}
\end{equation}
The spin-transfer torque is 
\begin{equation}
F_\Phi = - \frac{\partial E}{\partial \Phi}
    = - 2 \sigma \mathcal J u.
\label{eq:torque-without-glitches}
\end{equation}
This result is free from the glitches that plague Eq.~(\ref{eq:torque-with-glitches}).

\subsection{Vortex in two spatial dimensions}

For a rigidly moving vortex $\mathbf m(\mathbf r - \mathbf R)$, where $\mathbf R$ is the center of the vortex core,
\begin{equation}
\partial_s \mathbf m 
= \frac{\partial \mathbf m}{\partial X^i} \, \partial_s X^i
= - \partial_i \mathbf m \, \partial_s X^i.    
\end{equation}
The energy density in the extended 3-dimensional space $(x,y,s)$ is then 
\begin{equation}
T\indices{0}{0} 
    = \mathcal J u^i \mathbf m \cdot 
    (\partial_i \mathbf m \times \partial_j \mathbf m) \partial_s X^j
    = 4\pi \mathcal J u^i \rho \, 
    \epsilon_{ij} \partial_s X^j,
\end{equation}
where $\rho(x,y,s)$ is the 2-dimensional density of skyrmion charge (\ref{eq:skyrmion-density}). Integration over $x$, $y$, and $s$ yields the energy difference 
\begin{equation}
E_1 - E_0 =    4 \pi Q \mathcal J u^i  \epsilon_{ij} (X^j_1 - X^j_0) 
\end{equation}
and the force on the vortex
\begin{equation}
F^i = - \frac{\partial E}{\partial X^i}
= 4 \pi Q \mathcal J \epsilon_{ij} u^j,
\label{eq:F-vortex-take-3}
\end{equation}
in agreement with Eq.~(\ref{eq:F-vortex-take-2}).

The force can also be obtained directly from the stress tensor $T\indices{i}{j}$. The force density in the extended space $(\mathbf r,s)$ is  
\begin{eqnarray}
f^i 
    &=& \partial_j T\indices{i}{j} 
    = \mathcal J u^j \mathbf m \cdot 
    (\partial_i \mathbf m 
        \times \partial_s \partial_j \mathbf m
     - \partial_j \mathbf m 
        \times \partial_s \partial_i \mathbf m)
\nonumber
\\
    &=& \partial_s 
    [\mathcal J u^j \mathbf m \cdot
    (\partial_i \mathbf m 
        \times \partial_j \mathbf m)
    ],
\end{eqnarray}
where we used the identity $\partial_s \mathbf m \cdot (\partial_i \mathbf m \times \partial_j \mathbf m) = 0$, which holds because all three derivatives are orthogonal to $\mathbf m$ and are thus coplanar. Integration over the physical volume $V$ and then over the unphysical dimension yields the net force on the extended space: 
\begin{equation}
F^i_1 - F^i_0 
    = \int_0^1 ds \, \partial_s
    [4\pi \mathcal J Q \epsilon_{ij}u^j]
    = 4\pi \mathcal J (Q_1 - Q_0) \epsilon_{ij}u^j,
\end{equation}
in agreement with Eq.~(\ref{eq:F-vortex-take-3}).

\section{Are energy and momentum single-valued?}
\label{sec:single-valuedness}

The momentum difference between two magnetization configurations $\mathbf m_0(\mathbf r)$ and $\mathbf m_1(\mathbf r)$ is given by Eq.~(\ref{eq:Delta-P}). One may worry that the result will depend on the specific path between the two configurations. 

In a previous work \cite{Tchernyshyov2015}, one of us argued that conserved momenta of a ferromagnet are uniquely defined as long as the gyroscopic tensor satisfies the Jacobi identity. Here we shall present a different proof. In doing so, we will highlight the topological nature of the linear momentum. It remains unchanged under smooth deformations of the trajectory between configurations $\mathbf m_0(\mathbf r)$ and $\mathbf m_1(\mathbf r)$. However, the results may be different for topologically distinct paths, which cannot be smoothly deformed into each other. This latter point was not made in Ref.~\onlinecite{Tchernyshyov2015}.

A similar ambiguity may exist in the spin-torque part of the energy (\ref{eq:E-vortex}), $E^\text{ST} = \mathbf u \cdot \mathbf P$,
\begin{equation}
E^\text{ST}_1 - E^\text{ST}_0 = 
    - \mathcal J u^i \int dV \int_0^1 ds \, \mathbf m \cdot (\partial_i \mathbf m \times \partial_s \mathbf m).
\label{eq:E-EZ}
\end{equation}
We shall investigate the conditions under which the energy is single-valued. The analysis for linear momenta proceeds along the same lines and yields the same conclusions. 

To characterize different paths, it is convenient to parametrize the magnetization field $\mathbf m(\mathbf r)$ in terms of a (potentially infinite) set of collective coordinates $q(s) = \{q^a(s)\}$ so that $
\partial_s \mathbf m    
= (\partial \mathbf m/\partial q^a) \, dq^a/ds$. Then the energy difference can be written as an integral over these coordinates:
\begin{equation}
E^\text{ST}_1 - E^\text{ST}_0 
    = - \int_0^1 ds \, F_a \, \frac{dq^a}{ds}
    = - \int_{q(0)}^{q(1)} dq^a \, F_a,    
\label{eq:E-Fdq}
\end{equation}
where 
\begin{equation}
F_a = \mathcal J u^i \int dV 
    \mathbf m \cdot 
    \left(
        \partial_i \mathbf m
        \times 
        \frac{\partial \mathbf m}{\partial q^a}
    \right)
\label{eq:Fa}
\end{equation}
is a generalized force conjugate to coordinate $q^a$.

The energy difference (\ref{eq:E-Fdq}) will be a function of the endpoints $q(0)$ and $q(1)$ but not a functional of the entire path $q(s)$ if the integral along any closed loop in $q$-space vanishes. This statement is too strong and is violated in the case of a domain wall in one dimension, as discussed in Sec.~\ref{sec:T-gauge-invariant-domain-wall}.We will prove a weaker statement: the energy difference is invariant under smooth changes in the path from $q(0)$ to $q(1)$ and is thus the same for all homotopic (topologically equivalent) paths between $q(0)$ and $q(1)$. 

The energy difference (\ref{eq:E-Fdq}) is invariant under infinitesimal changes of the path from $q(0)$ to $q(1)$ if the force $F_a$ has zero curl,
\begin{eqnarray}
0 &=& \frac{\partial F_b}{\partial q^a} 
- \frac{\partial F_a}{\partial q^b}
\nonumber\\
&=& 
\mathcal J u^i \int dV 
    \mathbf m \cdot 
    \left[
        \frac{\partial \mathbf m}
            {\partial q^a}
        \times \partial_i 
        \frac{\partial \mathbf m}
            {\partial q^b}
        - (a \leftrightarrow b)
    \right]
\nonumber\\
&=& 
\mathcal J u^i \int dV \, 
    \partial_i 
    \left[
    \mathbf m \cdot 
    \left(
        \frac{\partial \mathbf m}
            {\partial q^a}
        \times 
        \frac{\partial \mathbf m}
            {\partial q^b}
    \right)
    \right].
\label{eq:0-curl}
\end{eqnarray}
By Stokes' theorem, the last line in Eq.~(\ref{eq:0-curl}) reduces to a surface integral over the spatial boundary of the system $\partial V$, thus giving the following condition:
\begin{equation}
\mathcal J u^i \int_{\partial V} dS_i \,
\mathbf m \cdot 
\left(
\frac{\partial \mathbf m}{\partial q^a}
\times
\frac{\partial \mathbf m}{\partial q^b}
\right)
= 0,
\label{eq:0-curl-surface}
\end{equation}
where $dS_i$ is the oriented surface element. In the simplest case, when the physical volume $V$ has no boundary, $\partial V = 0$, Eq.~(\ref{eq:0-curl-surface}) is automatically satisfied and thus the energy difference (\ref{eq:E-Fdq}) is insensitive to infinitesimal changes in the path from $q(0)$ to $q(1)$.

For a physical volume with a boundary, the integrand in Eq.~(\ref{eq:0-curl-surface}) can still vanish if variations of $\mathbf m$ under changes in collective coordinates do not affect magnetization at the sample boundary. Thus, if magnetization at the boundary is fixed or is confined to a single plane then the integrand vanishes, the force indeed has zero curl, and the energy is insensitive to continuous path changes. 

The requirement of magnetization remaining fixed or staying in a fixed plane at the boundary under arbitrary changes of collective coordinates may be a little too restrictive. We can relax it a bit by returning from the abstract space of collective coordinates to configurations of magnetization in spacetime $\mathbf m(t,\mathbf r)$. A path from $\mathbf m_0(t,\mathbf r)$ to $\mathbf m_1(t,\mathbf r)$ consists of consecutive increments $\delta \mathbf m$, $\delta \mathbf m'$, $\delta \mathbf m''$ etc. An infinitesimal change in the path can be made by switching two consecutive segments, e.g., $\delta \mathbf m'$ and $\delta \mathbf m''$. The corresponding change in the energy difference will be 
\begin{equation}
\mathcal J u^i \int_{\partial V} dS_i \,
\mathbf m \cdot 
(\delta \mathbf m'
\times
\delta \mathbf m''
)
\label{eq:dE-loop-config-space}
\end{equation}
So long as we stick to paths that keep magnetization at the boundary fixed or staying in the same plane, the energy will remain insensitive to the precise choice of the path. This works for both examples of topological defects considered in this paper, domain walls and vortices. In the latter case, vortex motion may affect magnetization at the boundary even if they remain far from the edge. However, the integrand in Eq.~(\ref{eq:dE-loop-config-space}) will vanish as long as vortex cores do not approach the boundary.

We see that, under fairly mild assumptions, the energy difference (\ref{eq:E-EZ}) is insensitive to smooth variations of the path between the $\mathbf m_0(\mathbf r,t)$ and $\mathbf m_1(\mathbf r,t)$. The same applies to the difference of momenta. 

\begin{figure}
    \includegraphics[width=0.95\columnwidth]{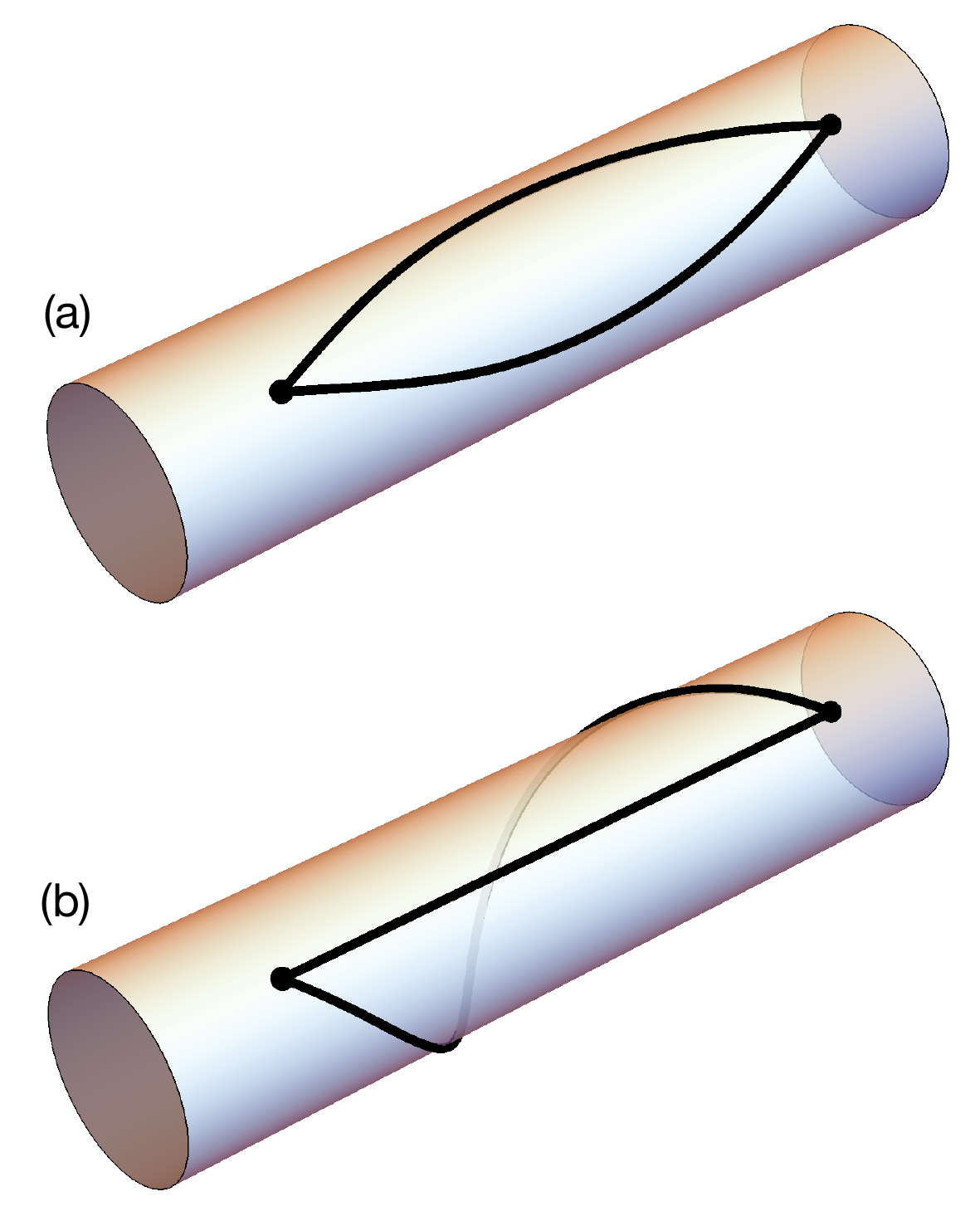}
    \caption{The space of collective coordinates of a one-dimensional domain wall restricted to its zero modes $q = \{X,\Phi\}$ is the surface of a cylinder, with $X$ along the cylinder axis and $\Phi$ along its circumference. Points symbolize initial and final configurations $q(0)$ and $q(1)$, lines represent paths $q(s)$ connecting them. (a) Homotopic paths. (b) Non-homotopic paths.}
    \label{fig:paths-on-cylinder}
\end{figure}

Note, however, that topologically distinct paths (those that cannot be continuously deformed into one another) can, in principle, yield different energy increments. The simplest example of that is a domain wall in one dimension discussed in Sec.~\ref{sec:T-gauge-invariant-domain-wall}. The $q$ space restricted to the zero modes $X$ and $\Phi$ is the surface of a cylinder, Fig,~\ref{fig:paths-on-cylinder}. The two paths in Fig.~\ref{fig:paths-on-cylinder}(a) are homotopic and thus yield the same energy difference (\ref{eq:E-EZ}). Those in Fig.~\ref{fig:paths-on-cylinder}(b) are not homotopic as they have increments $\Delta \Phi = 0$ and $2\pi$. As a result, the energy increments (\ref{eq:E-EZ}) along these paths differ by $4\pi \sigma \mathcal J u$, as can be seen from Eq.~(\ref{eq:E-domain-wall}).

This appears somewhat problematic because potential energy and momentum supposed to be a function of state and in this example it clearly is not. The ambiguity of the linear momentum can be resolved by appealing to quantum mechanics and to a discrete nature of the ferromagnet made of individual spins \cite{Haldane1986, Tchernyshyov2015}. The paradox with the energy can be resolved by extending the dynamical system to include the electric current flowing through the ferromagnet. From that standpoint, the spin-transfer torque is an instance of a gyroscopic force similar in nature to the Coriolis force in a rotating frame and to the Lorentz force on an electric charge in a magnetic field. Because gyroscopic forces do zero work, the energy of the full system, including the electric current, is single-valued \cite{Dasgupta-future}. 

\section{Discussion}

\label{sec:discussion}

The redefined energy-momentum tensor resolves paradoxes that arise when the canonical version of the tensor is used to calculate physical quantities such as momentum or energy. Gauge invariance is achieved at the cost of adding an extra space dimension, with the physical system living on its boundary. Examples discussed in Sec.~\ref{sec:T-gauge-invariant} show that the new gauge-invariant energy-momentum tensor provides a simple way to compute well-defined physical quantities. 

Integration of the energy-momentum tensor components $T\indices{\alpha}{0}$ over the physical volume and over the unphysical extra dimension yields the net energy and momentum of the extended spacetime. Only the boundaries of the extended spacetime, $s=0$ and 1, contribute to the net energy and momenta. The net energy is thus the difference $E_1 - E_0$ of the two ferromagnets living on the two boundaries. 

By construction, the magnetization configurations $\mathbf m_0(\mathbf r,t) = \mathbf m(\mathbf r,t,0)$ and $\mathbf m_1(\mathbf r,t) = \mathbf m(\mathbf r,t,1)$ are smoothly connected by an extension to the bulk of the extra dimension $\mathbf m(\mathbf r,t,s)$. A question arises then: do physical quantities such as energy and momentum depend on the precise trajectory from $\mathbf m(\mathbf r,t,0)$ to $\mathbf m(\mathbf r,t,1)$ along the extra dimension? 

The answer is two-fold. Infinitesimal changes of the trajectory do not influence the physical quantities. Thus all topologically equivalent paths from $\mathbf m(\mathbf r,t,0)$ to $\mathbf m(\mathbf r,t,1)$ yield the same energy and momentum difference between these configurations. However, topologically distinct paths may lead to different answers. An example of that is the energy of a domain wall in the presence of a spin current $E = \text{const} + 2\sigma \mathcal J u \Phi$. Rotating the plane of the domain wall by $\Delta \Phi = 2\pi$ increases the energy by $4\pi \sigma \mathcal J u$. Even though the domain wall returns to the same state, the path $\Delta \Phi = 2\pi$ is topologically distinct from the trivial path $\Delta \Phi = 0$ and the energy increments are different. This raises interesting questions since energy should be a function of state and in this case it appears not to be. A full resolution of this paradox requires a dynamical treatment of the electric current flowing through the domain wall. This topic lies outside of the scope of this paper and will be addressed in a forthcoming publication \cite{Dasgupta-future}. 

The energy-momentum tensor derived in this paper lives in an extended spacetime. Can we perhaps obtain a local energy-momentum tensor in the regular $(d+1)$-dimensional spacetime by integrating the $(d+2)$-dimensional version (\ref{eq:T-extra-dim}) over the extra dimension? The answer appears to be no. Take, for example, the energy density $T\indices{0}{0}$. By working along the lines of Sec.~(\ref{sec:single-valuedness}), we reproduce its results up to  Eq.~(\ref{eq:0-curl}), except we don't integrate over the physical volume $V$. Unfortunately, without the volume integration we cannot apply Stokes' theorem and thereby establish that the curl vanishes. Therefore, there is generally no way to obtain a local energy-momentum tensor in the physical $(d+1)$-dimensional spacetime. Thus our extension to $d+2$ dimensions is necessary if we wish the energy-momentum tensor to be a local quantity. 


\section*{Acknowledgments}

We thank Ibou Bah, Se Kwon Kim, and Shu Zhang for valuable discussions. This work was supported by the US Department of Energy, Office of Basic Energy Sciences, Division of Materials Sciences and Engineering under Award DE-FG02-08ER46544.

\appendix

\section{Wess-Zumino action for a single spin}
\label{app:WZ}

We briefly review the Wess-Zumino action for a single spin $\hbar \mathbf S$ of length $\hbar S$ \cite{Fradkin2013, Abanov2017}. The Landau-Lifshitz equation 
\begin{equation}
\hbar S \, \mathbf m \times \dot{\mathbf m} 
    - \frac{\partial U}{\partial \mathbf m}
    = 0    
\label{eq:LL-spin}
\end{equation}
for the unit vector $\mathbf m = \mathbf S/S$ can be derived from the standard action with a gauge-dependent potential $\mathbf a(\mathbf m)$,
\begin{equation}
S = \int dt 
    \left[
        \mathbf a(\mathbf m) 
        \cdot \dot{\mathbf m}
        - U(\mathbf m)
    \right],
\quad 
\nabla_{\mathbf m} \times \mathbf a 
= - \hbar S \mathbf m,
\end{equation}

Alternatively, we may replace the gauge-dependent term with a gauge-invariant Wess-Zumino action
\begin{equation}
S_\text{WZ} = \hbar S \int dt \int_0^1 ds \, 
    \mathbf m \cdot 
    (\partial_t \mathbf m 
    \times \partial_s \mathbf m).    
\label{eq:S-WZ-spin}
\end{equation}
The formal parameter $s$ can be thought of as an added spatial dimension, so that the spin lives in spacetime $(t,s)$ with $0 \leq s \leq 1$. Although the Wess-Zumino action is defined on the entire extended spacetime, its variation is limited to the boundaries $s=0$ and 1:
\begin{equation}
\delta S_\text{WZ} = \hbar S \int dt \,
(\mathbf m \times \partial_t \mathbf m) 
    \cdot 
    [\delta \mathbf m(t,1) 
    - \delta \mathbf m(t,0)].
\label{eq:dS-WZ-spin}
\end{equation}
The functional derivative $\delta S_\text{WZ}/\delta \mathbf m(t,1)$ yields the first term in the Landau-Lifshitz equation (\ref{eq:LL-spin}). Upon adding potential energy terms at the boundaries to the action,
\begin{equation}
S = S_\text{WZ} 
    - \int dt \, 
    [U(\mathbf m(t,1)) - U(\mathbf m(t,0))],
\label{eq:S-tot-spin}
\end{equation}
we obtain a system consisting of two independent spins $\mathbf m_0(t) = \mathbf m(t,0)$ and $\mathbf m_1(t) = \mathbf m(t,1)$ living at the opposite ends of the extra dimension $s=0$ and 1 and obeying the Landau-Lifshitz equation (\ref{eq:LL-spin}). 

The bulk of the spacetime $0 < s < 1$ is physically inert: the total action (\ref{eq:S-tot-spin}) is insensitive to how $\mathbf m(t,s)$ behaves there. Therefore all extensive physical quantities  (e.g., total energy and spin) characterizing the entire spacetime will be determined by what goes on at the boundaries. 

Note that the contributions from the two boundaries to Eqs.~(\ref{eq:dS-WZ-spin}) and (\ref{eq:S-tot-spin}) have opposite signs. Therefore extensive physical quantities characterizing the entire space $0 \leq s \leq 1$ will be obtained as the difference of the corresponding contributions from the two boundaries, rather than their sum. The alternating sign makes it possible to stack two spacetimes with $0 \leq s \leq 1$ and $1 \leq s \leq 2$ to form a seamless spacetime $0 \leq s \leq 2$. The actions from the two merged boundaries at $s=1$ cancel out.

As a simple example, let us compute the net angular momentum $\mathbf J$ of the extended system. The Wess-Zumino Lagrangian  
\begin{equation}
\mathcal L_\text{WZ} = 
    \hbar S \, \mathbf m \cdot 
    (\partial_t \mathbf m 
    \times \partial_s \mathbf m)    
\end{equation}
is invariant under global rotations, so Noether's theorem applies. An infinitesimal rotation through angle $\delta \phi$ about the unit vector $\mathbf n$ produces the variation $\delta \mathbf m = \mathbf n \times \mathbf m \, \delta \phi$. The corresponding component of the angular momentum is 
\begin{eqnarray}
\mathbf J \cdot \mathbf n &=& 
    \int_0^1 ds \, 
    \frac{\partial \mathcal L_\text{WZ}}
        {\partial \dot{\mathbf m}}
    \cdot 
    \frac{\partial \mathbf m}{\partial \phi}
\nonumber\\
    &=& \hbar S \int_0^1 ds \,
    (\partial_s \mathbf m \times \mathbf m)
    \cdot 
    (\mathbf n \times \mathbf m)
\\
    &=& \hbar S \int_0^1 ds \, 
    \partial_s \mathbf m \cdot \mathbf n
    = \hbar S \mathbf m_1 \cdot \mathbf n
    - \hbar S \mathbf m_0 \cdot \mathbf n.
\nonumber
\end{eqnarray}
As expected, the net angular momentum is the difference of the two spins, $\mathbf J = \hbar \mathbf S_1 - \hbar \mathbf S_0$. 

\bibliographystyle{apsrev4-1}
\bibliography{tensor}

\begin{thebibliography}{23}
\bibitem [{tra()}]{transverse}%
  \BibitemOpen
  \href@noop {} {}\bibinfo {note} {Keep in mind that $\mathbf m$ lives on a
  unit sphere and therefore its infinitesmial variation $\delta \mathbf m$ is
  tangential to the sphere and normal to $\mathbf m$. The functional derivative
  $-\delta U/\delta \mathbf m = M \mathbf H_\text{eff}$ is taken with respect
  to such transverse variations only, whereas the longitudinal component is
  undefined. Vector Eqs.~(\ref{eq:LLG}) and (\ref{eq:LLG-mod}) describe the
  motion tangential to the sphere and thus apply to components transverse to
  $\mathbf m$ only.}\BibitemShut {Stop}%
\end{thebibliography}

\end{document}